\documentstyle[psfig,prd,aps]{revtex}

\def\bbb{$b\overline{b} \,\,$}
\def\ccb{$c\overline{c} \,\,$}
\def\jpsi{$J/\psi$ \,\,}

\begin{document}
\draft
\title{Color Evaporation and Elastic
$\Upsilon$ Photoproduction at DESY HERA}
 \author{M.B. Gay Ducati
$^{1,\dag}$\footnotetext{$^{\dag}$ E-mail:gay@if.ufrgs.br}, V.P.
Gon\c{c}alves $^{2,*}$\footnotetext{$^{*}$
E-mail:barros@ufpel.tche.br}, C. B. Mariotto
$^{1,3,\star}$\footnotetext{$^{\star}$
E-mail:mariotto@if.ufrgs.br} }
\address{$^1$ Instituto de F\'{\i}sica, Universidade Federal do Rio Grande do
Sul\\
Caixa Postal 15051, CEP 91501-970, Porto Alegre, RS, BRAZIL}
\address{$^2$ Instituto de F\'{\i}sica e Matem\'atica,  Universidade
Federal de Pelotas\\
Caixa Postal 354, CEP 96010-090, Pelotas, RS, BRAZIL}
\address{$^3$ Department of Radiation Sciences,  Uppsala University, \\
Box 535, S-752 21 Uppsala, Sweden}
 \maketitle

\begin{abstract}
The diffractive photoproduction of vector mesons is usually
described considering the two-gluon (Pomeron)
 exchange, non-diagonal parton distributions and the contribution
 of the real part to the cross section.
In this letter we analyze the diffractive photoproduction of the
$\Upsilon$ at HERA using an alternative model, the Color
Evaporation Model (CEM), where the cross section is simply
determined by the boson-gluon cross section and an assumption for
the production of the colorless state. We verify that, similarly
to the $J/\Psi$ case, the HERA data for this process  can be well
described by the CEM. Moreover, we propose the analyzes of the
ratio $R = \sigma_{\Upsilon} / \sigma_{J/\Psi}$ to discriminate
between the distinct approaches.
\end{abstract}


The successful operation of the HERA $ep$ collider has opened a
new era of experimental and theoretical investigation into
diffractive vector-meson photo- and leptoproduction. On the
experimental side, the HERA accelerator extends the accessible
energy range by more than one order of magnitude over previous
experiments. The HERA data show that the cross sections for
exclusive vector-meson production rise strongly with energy when
compared to fixed-target experiments, if a hard scale is present
in the process. On the theoretical side, vector-meson production
has proven to be a very interesting process in which to test the
interplay between the perturbative and nonperturbative regimes of
QCD (For a review, see for example \cite {caldwell}). While in
inclusive process, as open heavy flavor production, the cross
section is described in terms of a perturbative term associated to
the cross section of the partonic subprocess, and a
nonperturbative term represented by the parton distributions, an
analogous factorization of hard and soft physics does not apply to
quarkonium production rates \cite{schu}, which constitute a small
fraction of the total open flavor cross section. This makes
quarkonium physics a challenging field \cite{schu}.

The current picture in the literature to describe the diffractive
photo- and leptoproduction of vector-mesons assumes that the color
singlet property of the meson is enforced at the perturbative
level by the two-gluon (Pomeron) exchange \cite{ryskin}. In this
approach the amplitude, in the target rest frame, is factorized as
a sequence of events very well separated
in time: (i) the photon fluctuates into a quark-antiquark pair; (ii) the $q%
\overline{q}$ pair scatters on the proton target, and (iii) the scattered $q%
\overline{q}$ pair turns into a vector meson. The interaction is
mediated by the exchange of two gluons in a color-singlet state.
Moreover, the two gluon exchange  amplitude can be shown to be
proportional to the gluon
distribution $xg(x,\overline{Q}^2)$, with $x=(M^2+Q^2)/W^2$ and $\overline{Q}%
^2\approx \frac 14(M^2+Q^2)$, where $W$ is the $\gamma p$ center
of mass energy and $M$ is the invariant mass of the
$q\overline{q}$ system. In the case of the production of a heavy
meson, the presence of the heavy meson mass ensures that
perturbative QCD can be applied even in  the  photoproduction
limit. This approach has been improved, particularly in the role
of the vector meson light cone wave function \cite {frastrik},
describing reasonably the HERA $J/\Psi $ data. When applied to the
recent HERA $\Upsilon $ data \cite{zeus,h1},  it only reproduces
the data if new effects, which significantly contribute, are
considered \cite{mrt,fms}: (a) the non-diagonal parton
distributions, which probe new nonperturbative information about
hadrons and are a generalization of the conventional parton
distributions (For a review see Ref. \cite{ji}); and (b) the real
part of the scattering amplitude. In Ref. \cite{fms}, a strong
correlation between the mass of the diffractively produced state
and the energy dependence of the total cross section was found,
implying a distinct energy dependence for the $\Upsilon $ and
$J/\Psi $ photoproduction. One of the main motivations for the
study of diffractive photoproduction of vector mesons in the
Pomeron model is the possibility of obtain a sensitive probe of
the behavior of the gluon distribution, due to the quadratic
dependence on the gluon distribution \cite{ryskin,nuc89} of the cross section in this
model.

An alternative view of the diffractive photoproduction process was
proposed recently  in Ref. \cite{cris}, where the $J/\Psi $
photoproduction was analyzed using the color evaporation model
(CEM) \cite {halzen}. In this case the color singlet is not
enforced at the perturbative level and the cross section for the
process is given essentially by the boson-gluon cross section plus
an assumption for formation of the colorless meson. In the CEM for
the diffractive photoproduction the cross section is linearly
proportional to the gluon distribution. These authors have
obtained a  parameter free prediction which describes very well
the HERA $J/\Psi $
data.

In this letter we extend the application of the CEM for the recent
HERA $\Upsilon $ data and verify that, using the parameters determined in $%
\Upsilon $ hadroproduction, this model reasonably describes the
experimental data with no need to introduce a colorless (Pomeron)
exchange at the  perturbative level, as well as, non-diagonal
parton distributions or the contribution of the real part of the
scattering amplitude. Moreover, we propose the analyzes of the
ratio $R=\sigma _\Upsilon /\sigma _{J/\Psi }$ to discriminate
between the distinct approaches.

Let us start from a brief review of the Color Evaporation Model
(CEM). One of the main uncertainties in the quarkonium production
is related to the transition  from the colored state to the
colorless meson. Initially, the $q\overline{q}$ pair will in
general be in a color octet state. It subsequently neutralizes its
color and binds  into a physical resonance. Color neutralization
occurs by interaction with the surrounding color field. If we
enforce that a colorless object is already present at the
perturbative level then a hard interaction with the surrounding
color field is assumed, and a minimal of two gluons should be
exchanged with the proton (Pomeron models). However, if the
colorless object is produced at the nonperturbative level, then
there is not a minimal restriction in the number of gluons
exchanged with the proton. The CEM provides a simple and general
phenomenological approach to color neutralization
(see also Ref. \cite{ingel}%
). In CEM, quarkonium production is treated identically to open
heavy quark production with exception that in the case of
quarkonium, the invariant mass of the heavy quark pair is
restricted to be below the open meson threshold, which is twice
the mass of the lowest meson mass  that can be formed with the
heavy quark. For bottomonium the upper limit on the
$b\overline{b}$ mass is then $2m_B$. Moreover, the CEM assumes
that the quarkonium dynamics is identical to all bottomonium
states, although the $b\overline{b}$ pairs are typically produced
at short distances in different color, angular momentum and spin
states. The hadronization of the bottomonium states from the
$b\overline{b}$ pairs is nonperturbative, usually involving the
emission of one or more soft gluons. Depending on the quantum
numbers of the initial $b\overline{b}$ pair and the final state
bottomonium, a different matrix element is needed for the
production of the bottomonium state. The average of these
nonperturbative matrix elements are combined into the universal
factor $F[nJ^{PC}]$, which is process- and kinematics-independent
and describes the probability that the $b\overline{b}$ pair binds
to form a quarkonium $\Upsilon (nJ^{PC})$ of given spin $J$,
parity $P$, and a charge conjugation $C$. Once $F$ has been fixed
for each state ($\Upsilon $, $\Upsilon ^{\prime }$ or $\Upsilon
^{\prime \prime }$) the model successfully predicts the energy and
momentum dependence \cite{vogt,satz}.

Considering the elastic $\Upsilon$ photoproduction at HERA, the
CEM predicts that the cross section is given by
\begin{eqnarray}
\sigma[\Upsilon(nJ^{PC})] = F[nJ^{PC}] \,\,\overline{\sigma}[b
\overline{b}] \,\,,
\end{eqnarray}
where the large distance factors $F$ can be written in terms of the probability
$1/9$ to
have a color singlet pair after the soft interactions, times the
fractions $\rho_i$ of total bottomonium carried by the different states, in a
similar way
as considered in \cite{halzen}, where the corresponding $\rho$ factors for charmonium
are claimed to be universal.
The short distance contribution is
\begin{eqnarray}
\overline{\sigma}[b \overline{b}] = \int_{2m_b}^{2m_B}
dM_{b\overline{b}}
\,\,\frac{d\sigma[b\overline{b}]}{dM_{b\overline{b}}} \,\,.
\label{cross}
\end{eqnarray}
Here $\sigma[b\overline{b}]$ is the spin- and color-averaged
cross section for
open
heavy quark production, which describes the boson-gluon fusion
process $\gamma g \rightarrow b \overline{b}$; $M_{b\overline{b}}$
is the invariant mass of
the $b\overline{b}$ pair, $m_b$ is the bottom quark mass and $2m_B$ is the $B%
\overline{B}$ threshold. The differential cross section for the
boson-gluon subprocess is well known (see Eq. (4) in Ref.
\cite{cris} for the expression at leading order). This will contribute to
elastic photoproduction, since in LO all energy of the photon is transferred to
the \bbb pair. Higher order corrections will be mostly important for the
inclusive and
the inelastic cross section.
Thus, the soft interactions have
a double
effect - to eliminate the color of the \ccb pair, allowing quarkonium
production, and to allow elastic production with a single perturbative gluon
exchange, by having soft gluon exchanges taking place in the non-perturbative
part of the process.

In CEM, the effect of these soft interactions is implicit in the
non-perturbative factors. An explicit modeling for the soft gluon exchanges is
done in the SCI model \cite{ingel}, which is based essentially in the same
general idea. In that model, by requiring either rapidity gaps or
leading proton in the final state, it was obtained explicitly a good
description of several diffractive processes, including diffractive \jpsi and
open beauty production in the Tevatron \cite{eitupps}. This shows that soft
gluon exchanges really might play a role in diffraction.

The CEM has been used to predict the production of $\Upsilon $ in
hadron and nuclear processes \cite{satz}, with the parameter
$F[nJ^{PC}]$ determined from the quarkonium production
experimental data at fixed-target energies. A remarkable feature
of the CEM is that this model also accounts (within $K$ factors)
for the quarkonium production at the Tevatron \cite{vogt}. As the
fixed target $%
\Upsilon $ data have generally given the sum of $\Upsilon $,
$\Upsilon ^{\prime }$ and $\Upsilon ^{\prime \prime }$ production,
due to the low mass resolution to clearly separate the peaks, only
the global factor $F[\Upsilon +\Upsilon ^{\prime }+\Upsilon
^{\prime \prime }]=0.044$ has been determined. This is an
effective value which reflects both direct production and chain
decays of higher  mass states. As discussed in Ref. \cite{gunion},
isolation of the direct production cross section for each
$\Upsilon _i$ requires the detection of the radiated photons
associated with chain decays, which is not currently available but
might be possible at the LHC. Considering some assumptions related
to the decay chain the following values for $F$ in direct
$\Upsilon _i$ production  have been estimated \cite{gunion}:
$F^d[\Upsilon ]=0.023$, $F^d[\Upsilon ^{\prime
}]=0.02$, $F^d[\Upsilon ^{\prime \prime }]=0.0074$,
where the $d$ superscript indicates the $F$ for direct production. In our
formalism the $F$ factors can be related to the $\rho$ factors by the simple
relation $F_i=1/9\rho_i$, where the $1/9$ is the probability to have the \bbb
pair in a color singlet state after the soft gluon exchanges, and $\rho_i$ are
the universal factors with give the fractions of the onium cross section
carried by the different onium states. Their corresponding values are, from
above,
\begin{eqnarray}
\rho^d[\Upsilon ]=0.207\,\,,\,\,\rho^d[\Upsilon ^{\prime
}]=0.18\,\,,\,\,\rho^d[\Upsilon ^{\prime \prime }]=0.066\,\,.
\end{eqnarray}
Once the free parameters have been determined in the hadronic
processes, we can use the CEM to predict the $\Upsilon $
photoproduction at HERA. This also follows the assumption used in \cite{zeus}
that the production ratios of $\Upsilon$, $\Upsilon ^{\prime}$ and
$\Upsilon ^{\prime \prime }$ are the same as those measured in hadron-hadron
collisions.
 A comment is in order here. In Ref.
\cite{gunion} the $\Upsilon $ hadroproduction has been calculated
considering the next-to-leading order contributions for the cross
section, which implies that the factor $F$ determined from  data
does not contain perturbative contributions  beyond leading order
and can be considered an universal factor which describes the
probability  for quarkonium production. Moreover, the calculations
in Ref. \cite{gunion} have used the MRS D$^{-^{\prime }}$ parton
densities as  input, but as
demonstrated in \cite{vogtcms}, where an  update from the analyzes of the $%
\Upsilon $ suppression for the CMS detector was made, the
$\Upsilon $ data can be equally well described using the GRV 94 LO
\cite{grv95} parton densities.

\begin{figure}[tbp]
\centerline{\psfig{file=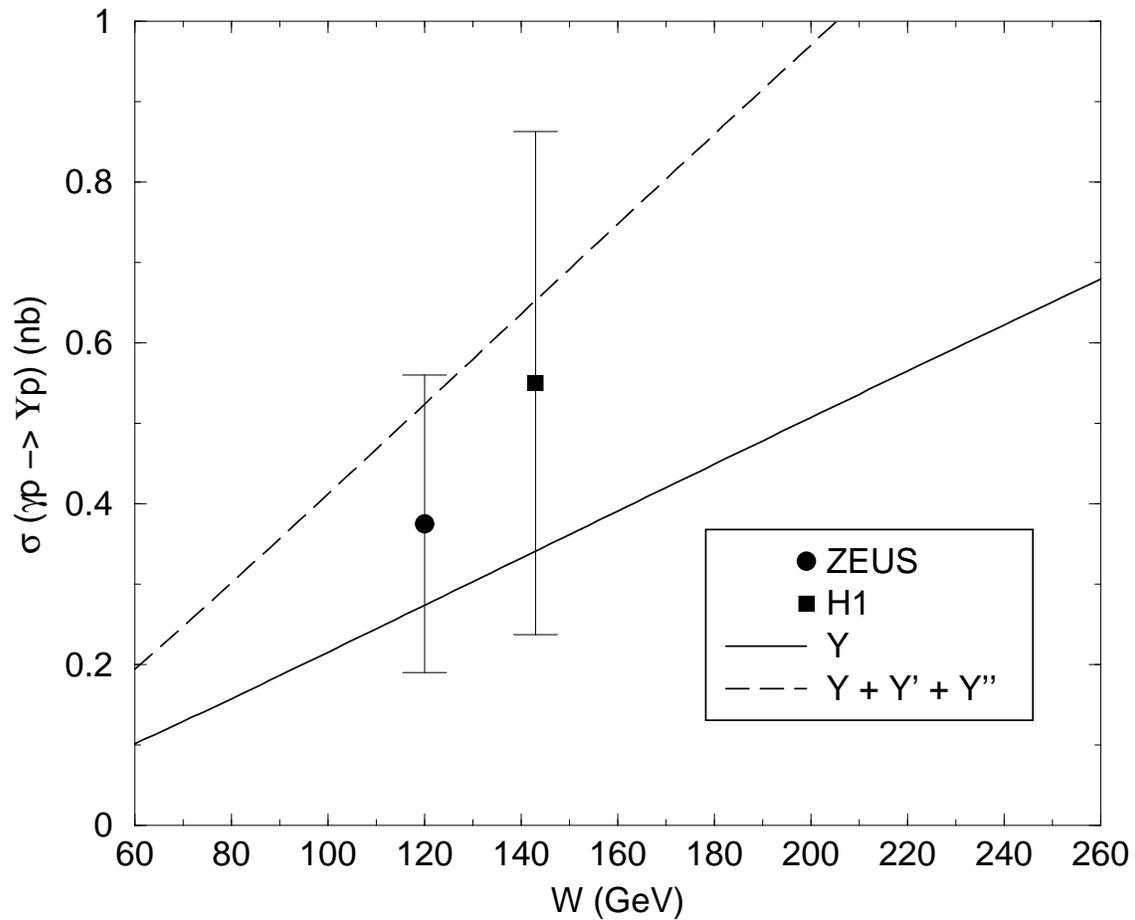,width=150mm}}
 \caption{Predictions
of the Color Evaporation Model for $\Upsilon$ photoproduction at
HERA. Cross section as a function of the photon-proton
center-of-mass energy $W$. The data points are the ZEUS  and H1
results for the direct production of the $\Upsilon$ state, which
should be compared with the full line. The total $\Upsilon
+\Upsilon '+\Upsilon ''$ contribution is also shown (dashed
line).}
 \label{fig1}
\end{figure}

The cross section $\overline{\sigma }[b\overline{b}]$ is computed
at leading order using the GRV 94 LO \cite{grv95} parton densities
with $m_b=4.75$ GeV, $m_B=5.279$ GeV and the renormalization and
factorization scales set to $\mu =(M_{Q\overline{Q}})^{1/2}$
\cite{halzen,cris}, with $Q = b$.  In Fig. \ref{fig1} we show our predictions for $%
\Upsilon $ photoproduction at HERA energies.  Both the total
$\Upsilon +\Upsilon ^{\prime }+\Upsilon ^{\prime \prime }$ and
direct $\Upsilon $ production are presented, since the cross
sections measured by the ZEUS and H1 collaborations did not select
the direct $\Upsilon $ state, but rather the data were integrated
over an interval of the $\mu ^{+}\mu ^{-}$ mass which includes at
least the $\Upsilon $, $\Upsilon ^{\prime }$ and $\Upsilon
^{\prime \prime }$ resonances. We verify that the current
experimental data does not allow to discriminate the distinct
contributions. We emphasize
that our results are completely parameter free  and that the CEM
model reasonably describes the scarce experimental data.

The simplicity of the CEM strongly contrasts with the number of
assumptions necessary in the Pomeron models to describe the same
set of data. Here we proposed a signature to discriminate between
these models. As we have quoted above, the Pomeron models predict
a stronger growth in energy for the diffractive $\Upsilon $
photoproduction than in $J/\Psi $ photoproduction, due to the
strong correlation between the mass of the diffractively produced
state and the energy dependence of the cross section. In contrast,
in the CEM model the growth of the cross section is directly
determined by the gluon distribution $xg(x,\mu )$, where $\mu $ is
the factorization scale. Therefore, the energy dependence of the
ratio
\begin{eqnarray}
R_{CEM}=\frac{\sigma _\Upsilon }{{\sigma _{J/\Psi }}}\,\,\,,
\label{ratiocem}
\end{eqnarray}
can be used to discriminate between models.

\begin{figure}[tbp]
\centerline{\psfig{file=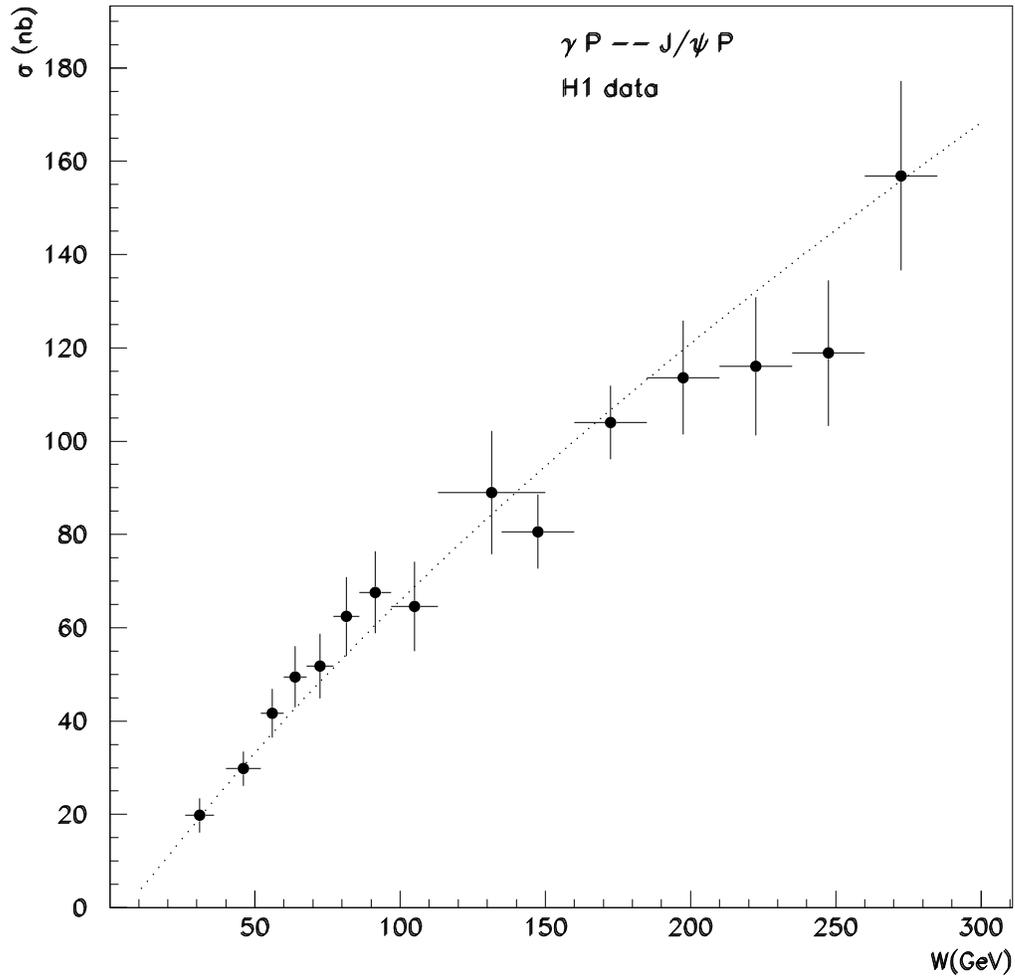,width=150mm}}
\caption{Predictions of the Color Evaporation Model for $J/\Psi$
photoproduction at HERA. Cross section as a function of the
photon-proton center-of-mass energy $W$. The data points are the
new H1  results for $J/\Psi$ production.} \label{fig3}
\end{figure}

 As an
intermediate step to calculate the mentioned rate, we show in Fig.
\ref{fig3} our previous calculation of $\sigma _{J/\Psi }$,
contrasted with the more recent HERA data \cite{h1}. We see that
these data can be reasonable explained without tuning the
parameters, which were taken as the previously used in Ref.
\cite{cris}. This gives one more clue that the CEM can be used to
explain elastic $J/\psi$ photoproduction.

\begin{figure}[tbp]
\centerline{\psfig{file=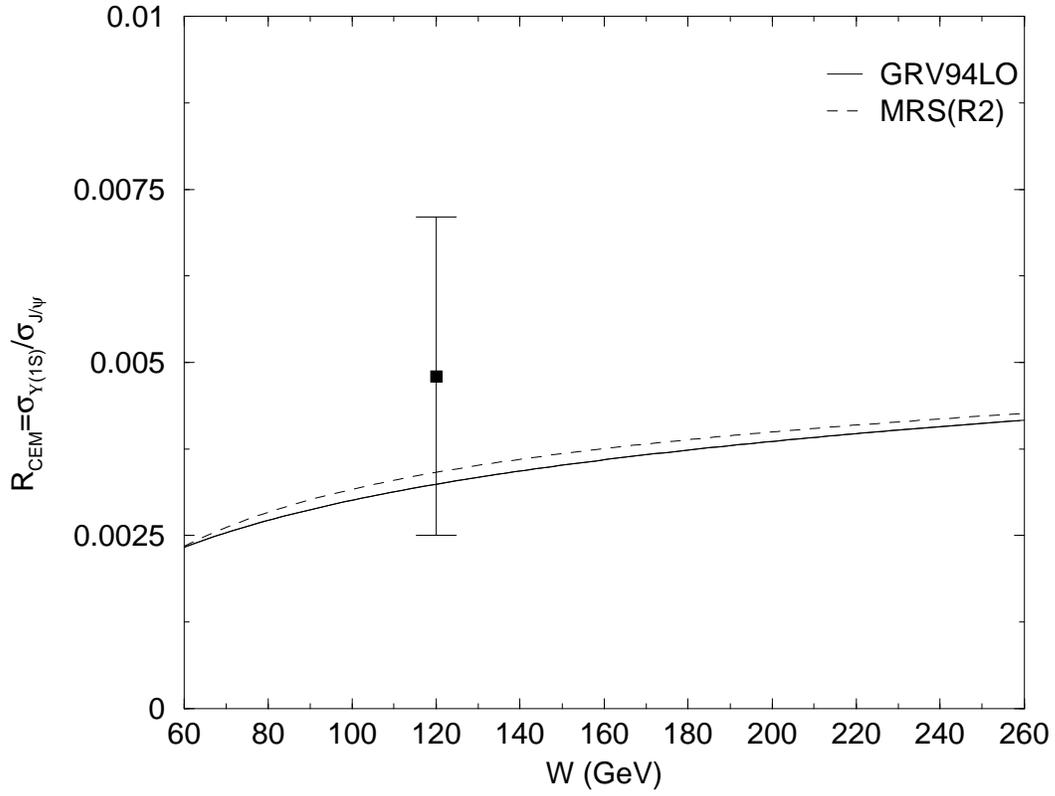,width=150mm}}
\caption{Predictions of the Color Evaporation Model for the energy
dependence of the ratio
$R_{CEM}=\frac{\sigma_{\Upsilon}}{{\sigma_{J/\Psi}}}$ in
photoproduction at HERA. Data from ZEUS.} \label{fig4}
\end{figure}

In Fig. \ref{fig4} we present the energy dependence of the ratio
$R_{CEM}={\sigma _\Upsilon }/{{\sigma _{J/\Psi }}}$ 
calculated using the CEM, where $\sigma _\Upsilon $ represents the direct $%
\Upsilon $ production and we have used the results from Ref.
\cite{cris} and from above to calculate $\sigma _{J/\Psi }$. Our
results show that this ratio is almost constant in the kinematic
range of HERA, in contrast to the Pomeron model
\cite{fms}, where a steep rise of the ratio is predicted [cf. Ref. \cite{fms}%
, $R_{Pom}\propto W$]. This ratio was also obtained in \cite{mrt},
where the value predicted agrees with data within errors for the
MRS gluon, but underestimates the result for the steeper GRV
gluon. In our case,
the $\sigma _\Upsilon /\sigma _{J/\Psi }$ ratio agrees with the
lower bound observed by ZEUS \cite{zeus} for both MRS and GRV
gluon parameterizations.

The original motivation of the study of diffractive
photoproduction was the possibility to extract the gluon
distribution inside the proton. However,  the dependence on the
gluon distribution is one of the major differences between the
descriptions of the diffractive photoproduction using either the
Pomeron model or the CEM. Whereas the Pomeron model has a
quadratic dependence on $xg$, this dependence is linear in the
CEM. Our result demonstrates that before to extract the gluon
distribution from HERA $\Upsilon$ and $J/\Psi$ data, one should
determine the correct description for this process, for example by
measuring the energy behaviour of the $\sigma _\Upsilon /\sigma
_{J/\Psi }$ ratio.

The CEM describes a large range of data in hadro- and
photoproduction, as shown in Refs. \cite{halzen,cris,cemmultip}.
Using this model, in this letter we obtain  a parameter free
description of the elastic $\Upsilon $ photoproduction at HERA
energies. We verify that this simple model reasonably describes
the experimental data, similarly to the Pomeron models. As a
distinct feature between these models, the CEM predicts a softer
energy dependence of the ratio between the $\Upsilon $ and $J/\Psi
$ cross sections. Of course, when more precise data become
available, the discrimination between these models should be
possible, which will allow us to conclude whether the CEM is only
 an alternative phenomenological model for the current energies
or it contains some underlying non-perturbative dynamics which is
important in both diffractive and non-diffractive quarkonium
production. In the latter case, more theoretical studies are
necessary to understand the soft interactions in the process of
color neutralization.

\section*{Acknowledgments}

This work was partially financed by CNPq, CAPES and FAPERGS,
BRAZIL.

\newpage

\end{document}